\documentclass[12pt]{article}
\usepackage{amsmath,amssymb,pictex,cite,eepic,epsfig,psfrag,url}
\usepackage{appendix}
\advance \topmargin by -\headheight
\advance \topmargin by -\headsep     
\evensidemargin \oddsidemargin
\makeatletter
\@addtoreset{equation}{section}

\makeatother
\def\Maketitle{{\def\newpage{}\maketitle}}
\begin{document}
\title{\textbf{A B\"acklund transformation for elliptic four-point conformal blocks\footnote{to be published in the Memorial Volume for Ludwig Faddeev.}}}
\author{Andr\'e Neveu\footnote{{\rm andre.neveu@umontpellier.fr}}\\
[1.6\bigskipamount]
~\parbox[t]{0.85\textwidth}{\normalsize\it\centering
L2C, CNRS, University of Montpellier, Montpellier, France}}

\date{}
\Maketitle
\begin{abstract}
We apply an integral transformation to solutions of a partial differential equation for five-point correlation functions in Liouville theory on a sphere with one degenerate field $V_{-\frac{1}{2b}}$. By repeating this transformation, we can reach a whole lattice of values for the conformal dimensions of the four other operators. Factorizing out the degenerate field leads to integral representations of the corresponding four-point conformal blocks. We illustrate this procedure on the elliptic conformal blocks discovered in a previous publication.
\end{abstract}
\section{Introduction}

Liouville field theory on a two-dimensional surface is defined by the Lagrangian density 
\begin{equation}\label{Lagrangian}
   \mathcal{L}=\frac{1}{4\pi}\hat{g}^{ab}\partial_a\varphi\partial_b\varphi+\mu {\rm e}^{2b\varphi}+
   \frac{Q}{4\pi}\hat{R}\,\varphi,
\end{equation}
where $\hat{g}_{ab}$ is the metric and $\hat{R}$ the associated curvature. In this paper, we shall exclusively consider the case of the sphere. The theory is conformally invariant if the coupling constant $b$ is related to the background charge $Q$ by
\begin{equation}\label{Q}
    Q=b+\frac{1}{b}.
\end{equation}
The central charge of the Virasoro algebra is 
\begin{equation}\label{c_L}
    c_L=1+6Q^2.
\end{equation}
The primary fields of the theory are associated with the exponential fields ${\rm{e}}^{2\alpha\varphi}$
and their conformal dimension is 
\begin{equation}\label{Delta}
   \Delta(\alpha)=\alpha(Q-\alpha).
\end{equation}
By conformal invariance, one can in principle find the correlation functions of all local fields from
the structure constants of the operator product expansion, starting from the structure constants of three primary fields $C_{\alpha_1\alpha_2}^{\alpha_3}$ which appear in the three-point correlation 
function\cite{Dorn:1992at,Dorn:1994xn,Teschner:1995yf,Zamolodchikov:1995aa}
\begin{equation}\label{3point-def}
  \langle V_{\alpha_1}(z_1)V_{\alpha_2}(z_2)V_{\alpha_3}(z_3)\rangle=\frac{C(\alpha_1,\alpha_2,\alpha_3)}
  {|z_{12}|^{2\Delta_1+2\Delta_2-2\Delta_3}|z_{13}|^{2\Delta_1+2\Delta_3-2\Delta_2}
  |z_{23}|^{2\Delta_2+2\Delta_3-2\Delta_1}},
\end{equation}
with 
\begin{equation}\label{structure}
    C_{\alpha_1\alpha_2}^{\alpha_3}=C(\alpha_1,\alpha_2,Q-\alpha_3).
\end{equation}
The four-point conformal block can then be represented as a sum over intermediate states
\begin{multline}\label{OPE-4point}
 \langle V_{\alpha_1}(z_1,\bar{z}_1)V_{\alpha_2}(z_2,\bar{z}_2)
         V_{\alpha_3}(z_3,\bar{z}_3)V_{\alpha_4}(z_4,\bar{z}_4)\rangle=
  \prod_{i<j}|z_i-z_j|^{2\gamma_{ij}}\\\times
  \frac{1}{2}\int\limits_{\mathcal{C}}C\Bigl(\alpha_1,\alpha_2,\frac{Q}{2}+{\rm i}P\Bigr)
  C\Bigl(\frac{Q}{2}-{\rm i}P,\alpha_3,\alpha_4\Bigr)
  \biggl|\mathfrak{F}_P
   \Bigl(\genfrac{}{}{0pt}{}{\alpha_2\;\;\alpha_3}{\alpha_1\;\;\alpha_4}\Bigl|x\Bigr)\biggr|^2
   dP,
\end{multline}
$x$ is the anharmonic ration of the four points $z_i$
\begin{equation}\label{crossratio}
  x=\frac{(z_1-z_2)(z_3-z_4)}{(z_1-z_3)(z_2-z_4)}.
\end{equation}
and the exponents $\gamma_{ij}$ which are standard combinations of the $\alpha$'s can be found in \cite{Belavin:1984vu}.

The conformal block  $\mathfrak{F}_P\Bigl(\genfrac{}{}{0pt}{}{\alpha_2\;\;\alpha_3}{\alpha_1\;\;\alpha_4}\Bigl|x\Bigr)$ sums up all the intermediate descendant states of a given primary one with conformal dimension $\Delta=\frac{Q^2}{4}-p^2$ in the operator product expansion \eqref{OPE-4point}.

In general, this function of $x$ is not known in closed form. However, when one of the external fields is degenerate it satisfies a linear differential equation \cite{Belavin:1984vu} which in the simplest case of the degenerate field $V_{-\frac{1}{2b}}$ can be solved explicitely in terms of hypergeometric functions.
In Ref.\cite{Fateev:2009jp} it was explained how to find more explicit expressions for conformal blocks
for $\alpha_4$, one of the parameters in Eq.\eqref{OPE-4point}, taking values of the form 
\begin{equation}\label{alpha4}
    \alpha_4=\frac{Q}{2}-\frac{b m_4}{2}+\frac{n_4}{b}-\frac{b}{4},
\end{equation}
for positive integers $m_4$ and $n_4$ with the other three $\alpha$'s equal to $Q/2-b/4$. In Section 2 we shall review this procedure, before explaining in Section 3 the extensions of these values of the $\alpha$'s by B\"acklund transformations. This paper can be considered a natural extension of Ref.\cite{Fateev:2009jp}, on which we rely heavily for notations and from which we reproduce several equations for completeness.

\section{Elliptic conformal blocks}

As in Ref.\cite{Fateev:2009jp}, let us consider the five-point correlation function with one degenerate field
\begin{equation}\label{5point}
\langle V_{-\frac{1}{2b}}(z)V_{\alpha_1}(0)V_{\alpha_2}(1)V_{\alpha_3}(\infty)V_{\alpha_4}(x)\rangle.
\end{equation}
Due to the presence of the degenerate field, this correlation function satisfies a partial differential equation, second order in $z$, first order in $x$. For the purposes of this section it is convenient to define the function $\Psi(u|q)$ as

\begin{equation}\label{5-point}
  \langle V_{-\frac{1}{2b}}(z)V_{\alpha_1}(0)V_{\alpha_2}(1)V_{\alpha_3}(\infty)
V_{\alpha_4}(x)\rangle
 = z^{\frac{1}{2b^2}}(z-1)^{\frac{1}{2b^2}}
  \frac{\bigl(z(z-1)(z-x)\bigr)^{\frac{1}{4}}}
  {\left(x(x-1)\right)^{\frac{2\Delta(\alpha_4)}{3}+\frac{1}{12}}}
  \frac{\Theta_1(u)^{b^{-2}}}{\Theta'_1(0)^{\frac{b^{-2}+1}{3}}}\,\Psi(u|q),
\end{equation}
where the variable $u$ is related with the variables $z$ and $x$ as
\begin{equation}\label{u-map}
    u=\frac{\pi}{4K(x)}\int_0^{\frac{z-x}{x(z-1)}}\frac{dt}{\sqrt{t(1-t)(1-xt)}},
\end{equation}
$K(x)$ is the elliptic integral of the first kind
\begin{equation}\label{ellipticK}
  K(x)=\frac{1}{2}\int_0^1\frac{dt}{\sqrt{t(1-t)(1-x t)}},
\end{equation} 
\begin{equation}\label{tau}
     \tau={\rm{i}}\,\frac{K(1-x)}{K(x)}=\frac{\log q}{{\rm{i}}\pi}, 
\end{equation}
and $\Theta_1(u)$ is the Jacobi theta function (see definitions in appendix). 

The function $\Psi(u|q)$  defined by Eq.\eqref{5-point} satisfies the non-stationary Schr\"odinger equation with doubly periodic potential
\begin{equation}\label{2order-diff-2}
   \Biggl[\partial^2_u-\mathbb{V}(u)+\frac{4{\rm i}}{\pi b^2}\partial_{\tau}\Biggr]\Psi(u|q)=0,
\end{equation}
where the potential $\mathbb{V}(u)$ is given by
\begin{equation}\label{double-periodic-potential}
  \mathbb{V}(u)=\sum_{j=1}^4s_j(s_j+1)\wp(u-\omega_j)
\end{equation}
and the parameters $s_k$ are related to the parameters $\alpha_k$ as
\begin{equation}\label{ak-sk}
    \alpha_k=\frac{Q}{2}-\frac{b}{2}\left(s_k+\frac{1}{2}\right).
\end{equation}
In Eq.\eqref{double-periodic-potential}  $\wp(u)$ is the Weierstra\ss  ~elliptic function with periods $\pi$ and $\pi\tau$ defined by the infinite sum
\begin{equation}\label{Wp}
  \wp(u)=\frac{1}{u^2}+\sum_{k^2+l^2\neq0}\left(\frac{1}{(u-\pi k-\pi\tau l)^2}-
  \frac{1}{(\pi k+\pi\tau l)^2}\right),
\end{equation}
and $\omega_j$ are the half periods
\begin{equation}\label{half-periods}
   \omega_1=\frac{\pi}{2},\qquad\omega_2=\frac{\pi\tau}{2},\qquad
   \omega_3=\omega_1+\omega_2,\qquad\omega_4=0.
\end{equation}

In Ref.\cite{Fateev:2009jp} explicit integral representations for the function $\Psi(u|q)$ 
are given for $s_1=s_2=s_3=0,\;s_4=m_4+\frac{2n_4}{b^2}$, $m_4$ and $n_4$ being integers.
For example, from the general solution to the heat equation:
\begin{equation}\label{Diff-double-periodic-free}
  \partial^2_u\Psi_0(u|q)+\frac{4{\rm i}}{\pi b^2}\partial_{\tau}\Psi_0(u|q)=0
\end{equation}
\begin{equation}\label{Psilibre}
    \Psi_0(u|q)=q^{\frac{b^2\lambda^2}{4}}{\rm e}^{-\lambda},    
\end{equation}
for $s_1=s_2=s_3=0$ and $s_4=1$, the general solution to the equation
\begin{equation}\label{Diff-double-periodic-1}
   \left(-\partial^2_u+2\wp(u)\right)\Psi(u|q)=\frac{4{\rm i}}{\pi b^2}\partial_{\tau}\Psi(u|q),
\end{equation}
can be obtained from the general solution $\Psi_0(u|q)$ of the  heat equation \eqref{Diff-double-periodic-free} as follows:
\begin{equation}\label{Lame-Sol-1}
  \Psi(u|q)=\int_0^{\pi}\left(\frac{\Theta_1(v)}{\Theta'_1(0)^{\frac{1}{3}}}\right)^{b^2}
  \frac{E(u+v)}{E(u)E(v)}\,\Psi_0(u+b^2v|q)\,dv,  
\end{equation}
where we introduced the notation
\begin{equation}\label{functionE}
    E(u)=\frac{\Theta_1(u)}{\Theta_1'(0)}.
\end{equation}

 In the limit of fixed $\lambda$ at $b\rightarrow\infty$ the integral in \eqref{Lame-Sol-1} is governed by the saddle point $\nu$, which is the solution of the equation
\begin{equation}\label{classical}
     \frac{\Theta_1'(\nu)}{\Theta_1(\nu)}=\lambda
\end{equation}
and $\Psi(u|q)$ has as limit (up to irrelevant normalization factors):
\begin{equation}\label{Stat-Lame-solution}
     \Psi(u|q)\rightarrow\frac{\Theta_1(u+\nu)}{\Theta_1(u)}{\rm e}^{-\lambda u},
\end{equation}
which is the solution of the stationary Lam\'e equation with energy $\wp(\nu)$. So, the solution \eqref{Lame-Sol-1} can be viewed as a ``quantization'' of the solution of the stationary equation 
\eqref{Stat-Lame-solution}.

In Ref.\cite{Fateev:2009jp} we explained how to obtain the four-point elliptic conformal blocks from the 
solutions $\Psi(u|q)$ by taking the limit $u\rightarrow 0$.

\section{B\"acklund transformations}

In this section, we find it more convenient to use the following parametrization for the 
five-point correlation function of Eq.\eqref{5-point}

\begin{multline}\label{5-pointnew}
<V_{\alpha_1}(0) V_{-\frac{1}{2b}}(z x) V_{\alpha_2}(x) V_{\alpha_3}(1) V_{\alpha_4}(\infty)>
=z^{(-\frac{1}{4b^2}+\frac{\alpha_1}{b}-\frac{Q}{2b})}x^{(\Delta-\Delta_1)}\\
\times
(1-z)^{\alpha_2/b}(1-x)^{-2\alpha_2\alpha_3}(1-z x)^{\alpha_3/b}g(z,x).
\end{multline}

This parametrization extracts the short distance behaviors at $z=1$, $x=1$, $z x=1$
which one obtains for the $V$'s being exponentials of free fields and taking for the dimension 
of the intermediate state between $V_{-\frac{1}{2b}}(z x)$ and $V_{\alpha_2}(x)$ the value
$(\alpha_1-\frac{1}{2b})(Q-\alpha_1+\frac{1}{2b})$. $\Delta=\alpha(Q-\alpha)=Q^2/4-p^2$ is the dimension, which should be integrated over, 
of the intermediate state between  $V_{\alpha_2}(x)$ and $V_{\alpha_3}(1)$. Then, $g(z,x)$ should 
be a series in $z$ and $x$ with only integer powers, whose coefficients can be 
computed recursively.

As in the previous section, one obtains the following partial differential equation for $g(z,x)$:

\begin{multline}\label{equ}
z(1-z)(1-z x){\partial_{z}}^2g(z,x)+(A+B z+C z x+D z^2x)\partial_{z}g(z,x)-\\
-x(1-x)\partial_{x}g/b^2 +(E+F x +H z x)g=0,
\end{multline}
with
\begin{equation}\label{polynomials}
  \begin{aligned}
    &A=2\alpha_1/b-1/b^2,\\
    &B=-2\alpha_1/b-2\alpha_2/b+2/b^2,\\
    &C=-2\alpha_1/b-2\alpha_3/b+1/b^2,\\
    &D=2\alpha_1/b+2\alpha_2/b+2\alpha_3/b-2/b^2,\\
    &E=\bigl(\alpha +\alpha_1 +\alpha_2 -1/(2b) -Q\bigr)\bigl(\alpha-\alpha_1-\alpha_2+1/(2b)\bigr)/b^2,\\
    &F=\bigl(-2\alpha_3(\alpha +\alpha_1 +\alpha_2 -1/(2b) -Q)-(\alpha_3+\alpha_4-\alpha)(\alpha_3-\alpha_4-\alpha+Q)\bigr)/b^2,\\
    &H=\bigl(\Delta_4-Q^2/4 +(\alpha_3+\alpha_2+\alpha_1-1/(2b)-Q/2)^2\bigr)/b^2.
  \end{aligned}
\end{equation}

Expanding $g(z,x)$ in powers of $x$:
\begin{equation}\label{gexpansion}
g(z,x)=g_0(z)+x g_1(z)+{x}^2 g_2(z)...,
\end{equation}
one obtains for $g_0$ a hypergeometric equation, and for the successive $g_i$'s
hypergeometric equations with non-zero right-hand sides which depend on the lower $g_i$'s.
For generic values of the $\alpha$'s and $b$, these equations have unique solutions with a power series expansion at $z=0$ and irrational power behaviors at $z=1$ and $z=\infty$.
The conformal block 

\begin{equation}\label{confblock}
<V_{\alpha_1-1/(2b)}(0) V_{\alpha_2}(x) V_{\alpha_3}(1) V_{\alpha_4}(\infty)>
\end{equation}
is then given by the function  $x^{(\Delta-\Delta_1')}(1-x)^{-2\alpha_2\alpha_3}g(0,x)$ 
with $\Delta_1'=(\alpha_1-\frac{1}{2b})(Q-\alpha_1+\frac{1}{2b})$. To study the
properties of this conformal block, for example its bootstrap properties,
one thus needs an explicit solution of Eq.\eqref{equ} to all orders in $x$.
Some non-trivial such solutions in terms of elliptic functions have been found 
in Ref.\cite{Fateev:2009jp}. In this section, we now wish to substantially expand this 
set of solutions.

Consider the integral transformation

\begin{equation}\label{trans}
g_I(z,x) = \int \frac{du}{(z-u)^y} \, g(u,x).
\end{equation}
This is inspired by formula (2) of subchapter (2.4) of Ref.\cite{Bateman:1953ht}
which transforms one hypergeometric function into another with
different values of the parameters. 
We leave the integration limits unspecified. The only requirement
is that boundary terms in partial integrations can be neglected:
For this, one may have a closed contour or end points where the integrand
exhibits exclusively non-integer power behaviors.

Look for values of $y$ such that $g_I$ is a solution of the 
same equation, but with new values of its parameters:

\begin{multline}\label{equ2}
z(1-z)(1-z x){\partial_{z}}^2g_I(z,x)
+(A_I+B_I z+C_I z x+D_I{z}^2 x)\partial_{z}g_I\\
-x(1-x)\partial_{x}g_I/b^2 +(E_I+F_I x +H_I z x)g_I=0.
\end{multline}

This is straightforward. One finds that $y$ must satisfy

$$ y^2 + y (1-D_I) + H_I = 0$$
together with the following relations with the initial coefficients:

\begin{equation}\label{newcoeffs}
\begin{aligned}
&A = A_I + 1-y\\
&B = B_I - 2(1-y) \\
&C = C_I - 2(1-y) \\
&D = D_I + 3(1-y) \\
&E = E_I + (y + B_I) (1-y) \\
&F = F_I + (y + C_I) (1-y) \\
&H = H_I + (2 D_I - 3y)(1-y).
\end{aligned}
\end{equation}

One can of course invert these relations, and see how the 
external $\alpha_i$'s are transformed into $\alpha_{I,i}$. One finds
(the signs are correlated, they correspond to the two solutions
for the quadratic equation for $x$, and obviously to the fact
that $\alpha_4$ appears in the original equation only through $\Delta_4$):

$$
y=1-1/(2b^2)+(-\alpha_1+Q/2-\alpha_2+Q/2-\alpha_3+Q/2.
\mp\alpha_4\pm Q/2)/b
$$

By analogy with string theory, introducing momenta $p_i$ for the vertex operators
through the definition $p_i=\alpha_i-Q/2$, this reads

$$
y=1-1/(2b^2)-(p_1+p_2+p_3\pm p_4)/b,
$$

\begin{equation}\label{newalphas}
\begin{aligned}
&\alpha_{I1}-Q/2=-1/(4b)+(\alpha_1-Q/2-\alpha_2+Q/2-\alpha_3+Q/2
\mp\alpha_4\pm Q/2)/2,\\
&\alpha_{I2}-Q/2=-1/(4b)+(-\alpha_1+Q/2+\alpha_2-Q/2-\alpha_3+Q/2
\mp\alpha_4\pm Q/2)/2,\\
&\alpha_{I3}-Q/2=-1/(4b)+(-\alpha_1+Q/2-\alpha_2+Q/2+\alpha_3-Q/2
\mp\alpha_4\pm Q/2)/2,\\
&\alpha_{I4}-Q/2=-1/(4b)+(-\alpha_1+Q/2-\alpha_2+Q/2-\alpha_3+Q/2
\pm\alpha_4\mp Q/2)/2.
\end{aligned}
\end{equation}

The intermediate dimension is unchanged:

$$
\alpha_I=\alpha.
$$

In the following, we shall take the upper signs in these equations,
simply remembering that the sign of $p_4$ is arbitary; changing it
is just choosing the other signs.

One introduces the matrix
\begin{equation}\label{matrixM}
M = \left( \begin{array}{rrrr}
\frac{1}{2} & -\frac{1}{2} & -\frac{1}{2} & -\frac{1}{2} \\ \\
-\frac{1}{2} & \frac{1}{2} & -\frac{1}{2} & -\frac{1}{2} \\ \\ 
-\frac{1}{2} & -\frac{1}{2} & \frac{1}{2} & -\frac{1}{2} \\ \\ 
-\frac{1}{2} & -\frac{1}{2} & -\frac{1}{2} & \frac{1}{2}
\end{array} \right)
\end{equation}
which is a symmetry with respect to the hyperplane perpendicular 
to the vector $(1,1,1,1)$ in the four-dimensional space of the
values of the $p_i=\alpha_i-Q/2$. The integral transformation from 
$g$ to $g_I$ is thus a reflection with respect to that hyperplane, 
followed by a translation by $-1/(4b)$ along the vector  $(1,1,1,1)$.

Now, one can change the signs of the $p_i=\alpha_i-Q/2$, $i=1,2,3$:
By the change of function  $g=z^{(Q-2\alpha_1)/b}g_1$ one finds that 
$g_1$ satisfies Eq.\eqref{equ} with $p_1$ changed into $-p_1$, $p_2$,
$p_3$,  $\Delta_4$ and $p$ unchanged; in the same fashion, by the change 
 $g=((1-z)/(1-z x))^{(Q-2\alpha_2)/b}g_2$ one finds that 
$g_2$ satisfies Eq.\eqref{equ} with $p_2$ changed into $-p_2$, $p_1$,
$p_3$,  $\Delta_4$ and $p$ unchanged, and finally the change 
$g=(1-z x)^{(Q-2\alpha_3)/b}g_3$ changes the sign of $p_3$. These 
changes of signs correspond to symmetries with respect to the 
coordinate hyperplanes.

By successive applications of these geometrical transformations, one generates 
a lattice in the space of the values of the $\alpha_i-Q/2$.
Calling $R_i$ the change of sign of $p_i=\alpha_i-Q/2$, and $I$
the integral transformation above, the sequence $R_1 \cdot I \cdot R_2 \cdot R_3 \cdot R_4
\cdot I \cdot R_2 \cdot R_3 \cdot R_4 \cdot I$ transforms any vector

\begin{equation}\label{K_0}
K_0 =  \left( \begin{array}{r}
p_1\\ \\ 
p_2\\ \\ 
p_3\\ \\ 
p_4 
\end{array} \right) \,\,\,\,{\rm into}\,\,\,\,
K_1 =  
\left( \begin{array}{c}
p_1 + 1/b  \\ \\ 
p_2 \\ \\ 
p_3 \\ \\ 
p_4
\end{array} \right)
\end{equation} 

By repeating this set of manipulations, one can clearly in principle reach an arbitrary
point of the lattice:
\begin{equation}\label{lattice}
K_0+ (n_1 e_{01} +
n_2 e_{02} + n_3 e_{03} + n_4 e_{04})/b
\end{equation}

where the vectors $e_{0i}$ are defined by:

\begin{equation}\label{basevectors}
e_{01}=\left( \begin{array}{r}
1 \\ 0\\ 0\\ 0
\end{array} \right) \,\,\,
e_{02}=\left( \begin{array}{r}
0 \\ 1\\ 0\\ 0
\end{array} \right) \,\,\,
e_{03}=\left( \begin{array}{r}
0 \\ 0\\ 1\\ 0
\end{array} \right) \,\,\,
e_{04}=\left( \begin{array}{r}
0 \\ 0\\ 0\\1
\end{array} \right)
\end{equation}

and the $n_i$ are arbitrary positive or negative integers. This partly answers
the conjecture put forward in \cite{Fateev:2009jp} on the values of the external
dimensions for which an explicit representation could be obtained.

By applying only once the integral transform on $K_0$, one generates 
the solutions for the vector

\begin{equation}\label{K_2}
K_2 =  \left( \begin{array}{c}
(p_1-p_2-p_3-p_4)/2-1/(4b)\\ \\ 
(-p_1+p_2-p_3-p_4)/2-1/(4b)\\ \\ 
(-p_1-p_2+p_3-p_4)/2-1/(4b)\\ \\ 
(-p_1-p_2-p_3+p_4)/2 -1/(4b)
\end{array} \right) 
\end{equation} 
which is another starting point from which one can reach the points 
of another lattice as for $K_0$ above.

Still another starting point for the construction of a lattice is 
the vector $K_3$ obtained by applying the sequence $ I  \cdot R_3 \cdot R_4 \cdot I$
to $K_0$. One finds

\begin{equation}\label{K_3}
K_3 =  \left( \begin{array}{c}
-p_2-1/(2b)\\ \\ 
-p_1-1/(2b)\\ \\ 
p_4\\ \\ 
p_3 
\end{array} \right) 
\end{equation}

Choosing the values $p_i=-b/4$ in $K_2$, applying the integral transformation, then the $R_1$
symmetry, one obtains the solution of Eq.\ref{equ2} for the following vector of momenta

\begin{equation}\label{K_4}
K_4 =  \left( \begin{array}{c}
-b/4+1/(4b)\\ \\ 
b/4-1/(4b)\\ \\ 
b/4-1/(4b)\\ \\ 
b/4-1/(4b)
\end{array} \right) 
\end{equation} 

Then, the value $p_1'$ which appears in the conformal block \eqref{confblock} is $-b/4-1/(4b)=-Q/4$.
Together with the value of the other $p_i$, $b/4-1/(4b)$, we thus obtain an explicit integral 
representation of a conformal block for $0<Q<4$ ($1<c_L<25$) where $b$ is unimodular, $p_1'$ real, 
the other $p_i$ pure imaginary, all conformal dimensions of external operators real and positive and
the dimension $\Delta$ of the intermediate state arbitrary. If we start from the opposite vector
$p_i=b/4$, the same transformation gives us $p_1'=b/4-1/(4b)$, pure imaginary for $1<c_L<25$ 
while the three other $p_i=-b/4-1/(4b)=-Q/4$ are real for the same values of $c_L$. As far as we know,
these two cases are the first explicit examples of four-point conformal blocks with positive dimensions at $1<c_L<25$ where the bootstrap could be explored analytically
as was done in \cite{Fateev:2009jp}, but this would go beyond the scope of this paper.

\section{Conclusion}

By applying integral transformations we have extended the cases of Ref.\cite{Fateev:2009jp} where explicit expressions for 
four-point conformal blocks could be found in terms of integrals involving elliptic funtions. 
There remains the problem of obtaining still further explicit representations in the cases
where the parameters $s_k$ of Eq.\eqref{ak-sk} take arbitrary integer values. 
For the moment, 
such more general expressions are only known in the ``classical'' case $b\rightarrow\infty$.
From Ref.\cite{Fateev:2009jp} we only have explicit expressions 
at finite $b$ for the cases of only one $s_k$
non-vanishing, or, via duplication formulas for elliptic functions, two of them equal
with the other two vanishing, or, still, all four equal. 

\section*{Appendix}

The theta function $\Theta_1(u|\tau)$ defined by
\begin{equation*}
    \Theta_1(u|\tau)=\sum_{n=-\infty}^{\infty}(-1)^{n-\frac{1}{2}}q^{(n+\frac{1}{2})^2}{\rm e}^{(2n+1){\rm i}u}  
\end{equation*}
is solution to the differential equation
\begin{equation*}
   \partial_u^2\Theta_1(u|\tau)-\frac{4{\rm i}}{\pi}\partial_{\tau}\Theta_1(u|\tau)=0.
\end{equation*}
It satisfies the following quasi-periodicity relations 
\begin{equation*}
  \begin{aligned}  
   &\Theta_1(u+\pi)=-\Theta_1(u),\\
   &\Theta_1(u+\pi\tau)=-q^{-1}{\rm e}^{-2{\rm i}u}\Theta_1(u),  
  \end{aligned}  
\end{equation*}
and transforms as follows under the action of the modular group
\begin{equation*}
   \begin{aligned}  
   &\Theta_1(u|\tau+1)={\rm i}^{\frac{1}{2}}\Theta_1(u|\tau),\\
   &\Theta_1\left(\frac{u}{\tau}\Bigl|-\frac{1}{\tau}\right)=({\rm i}\tau)^{\frac{1}{2}}{\rm e}^{\frac{{\rm i}u^2}{\pi\tau}}
   \Theta_1(u|\tau).  
  \end{aligned}
\end{equation*}
Other theta-functions can be expressed through $\Theta_1(u)$ as
\begin{equation*}
   \begin{aligned} 
   &\Theta_2(u)=\Theta_1\left(u+\frac{\pi}{2}\right),\\
   &\Theta_3(u)={\rm e}^{{\rm i}u}q^{\frac{1}{4}}\Theta_1\left(u+\frac{\pi}{2}+\frac{\pi\tau}{2}\right),\\
   &\Theta_4(u)=-{\rm i}{\rm e}^{{\rm i}u}q^{\frac{1}{4}}\Theta_1\left(u+\frac{\pi\tau}{2}\right).
  \end{aligned}
\end{equation*}

\bibliographystyle{MyStyle} 
\bibliography{MyBib}
\end{document}